\documentclass{article}

\usepackage{booktabs}
\usepackage{array}
\usepackage{graphicx}
\usepackage{amsmath}
\usepackage{amsfonts}
\usepackage{amssymb}
\usepackage{multirow}
\usepackage{url}

\begin{document}

\title{A hypothesis test of feasibility for external pilot trials assessing recruitment, follow-up and adherence rates}

\author{Duncan T. Wilson \and
	Rebecca E. A. Walwyn \and
	Julia Brown \and
	Amanda J. Farrin }
\date{Leeds Institute of Clinical Trials Research, University of Leeds, Leeds, UK (d.t.wilson@leeds.ac.uk)}

\maketitle

\begin{abstract}
The power of a large clinical trial can be adversely affected by low recruitment, follow-up and adherence rates. External pilot trials estimate these rates and use them, via pre-specified decision rules, to determine if the definitive trial is feasible and should go ahead. There is little methodological research underpinning how these decision rules, or the sample size of the pilot, should be chosen. In this paper we propose a hypothesis test of the feasibility of a definitive trial, to be applied to the external pilot data and used to make progression decisions. We quantify feasibility by the power of the planned trial, as a function of recruitment, follow-up and adherence rates. We use this measure to define hypotheses to test in the pilot, propose a test statistic, and show how the error rates of this test can be calculated for the common scenario of a two-arm parallel group definitive trial with a single normally distributed primary endpoint. We use our method to re-design TIGA-CUB, an external pilot trial comparing a psychotherapy with treatment as usual for children with conduct disorders. We then extend our formulation to include using the pilot data to estimate the standard deviation of the primary endpoint. and incorporate this into the progression decision.
\end{abstract}

\section{Introduction}\label{sec:intro}

Randomised Controlled Trials (RCTs) often fail to recruit to target \cite{Sully2013}, leading to an analysis with less power than intended. Power can also be adversely affected by large rates of attrition, poor adherence to the allocated treatment \cite{Fay2006}, and incorrect specification of a nuisance parameter such as the standard deviation of a continuous endpoint \cite{Teare2014}. A common approach to anticipate these problems is to run a small version of the intended trial, known as an external pilot, to obtain estimates of the relevant parameters and decide if, and how, the definitive trial should be conducted  \cite{Craig2008, Eldridge2016}. This decision is typically made with reference to so-called \emph{Progression Criteria} (PCs), a set of conditions which must be satisfied for the definitive trial to be considered feasible \cite{Avery2017}. In the case of quantitative PCs, parameter estimates are computed using the pilot data and then compared against threshold values. If all estimates exceed their respective thresholds, it is recommended that the definitive trial should be conducted. A recent workshop identified that recruitment, follow-up, and intervention adherence have emerged as common targets of progression criteria \cite{Avery2017}. 

The key role of PCs in pilot trials has led to the CONSORT extension for randomised pilots requiring their reporting \cite{Eldridge2016a}, and the NIHR, one of the major funders of pilot trials in the UK, requiring their pre-specification in the research plan \cite{NIHR2017}. There are ethical and economic imperatives to ensuring PCs are appropriate. If they are too lenient, we increase the risk of proceeding to the definitive trial only to discover it is infeasible, fail to answer the scientific question, in the process wasting resources and subjecting patients to research unnecessarily. If they are too strict, we increase the risk of failing to conduct a definitive trial which would in fact have been feasible, and as a result withholding a promising intervention from patients. Despite all this, there is little methodological research available to help researchers determine optimal PCs \cite{Avery2017}. 



Another important aspect of external pilot trial design is the choice of sample size. Here, methods have generally focussed on obtaining a sufficiently precise estimate of the standard deviation of a continuous primary outcome to inform the sample size calculation of the definitive trial \cite{Teare2014, Browne1995, Julious2005, Sim2012, Eldridge2015, Whitehead2015}. These methods, often reduced to simple rules-of-thumb such as requiring 35 participants in each arm \cite{Teare2014}, are nevertheless widely used to choose pilot sample size when the estimation of the standard deviation is not the only, or primary, objective. When the goal of the pilot is to provide estimates of recruitment, follow-up and adherence rates and use these in PCs, imprecise estimates will be more likely to meet or miss a PC threshold by chance alone \cite{Eldridge2015, Cooper2018} and thus lead to incorrect progression decisions. This issue will be compounded whenever there are several PCs, with progression to the definitive trial permitted only if all are satisfied. As the number of PCs grows, so does the probability of missing at least one threshold through bad luck, even when the precision around each individual estimate appears adequate - the so-called `reverse-multiplicity' problem seen in multiple testing procedures \cite{Senn2007, Chuang-Stein2007}. Although the sample size of pilots is often justified by reporting the anticipated precision in the estimates of feasibility parameters (e.g. the widths of confidence intervals), there is no clear guidance on how precise they should be, or how to weigh precision against the cost of sampling.



A more formal approach to the design and analysis of external pilot trials could employ a hypothesis testing framework. Taking this view, progression to the definitive trial would be determined by comparing an appropriate statistic to a critical value rather than using several independent PCs. To implement this approach we must be able to  identify levels of recruitment, follow-up and adherence which would correspond to feasible or infeasible definitive trials, and identify an approproiate statistic that can differentiate between them. Given these, the critical value and the pilot sample size can then be chosen with regards to the long-run error rates they lead to. 
Although many authors have advised against conducting formal hypothesis tests in pilot trials \cite{Lancaster2004, Arain2010, Thabane2010, Eldridge2015}, these warnings have been in the context of tests of effectiveness. Assuming the effect size of interest is the same in the pilot and the definitive trial and that conventional type I error rates are used, such a test will have low power. Tests assessing rates of recruitment, follow-up and adherence will not necessarily suffer from the same restriction. Moreover, it should be emphasised that the conventional method of pre-specifying several PCs and using these to map pilot data to progression decisions is mathematically equivalent to a multivariate hypothesis test, but without hypotheses being defined and therefore without any understanding of the statistical properties of the resulting procedure. A formal approach is therefore of interest, both to gain an understanding of current practice and to investigate if, and how, progression decisions can be improved. 



The remainder of the paper is structured as follows. First, we will define the specific problem under consideration in Section \ref{sec:problem}. In Section \ref{sec:methods} we will describe a formal hypothesis test of feasibility based on recruitment, follow-up and adherence rates. We will show how null and alternative hypotheses can be defined in terms of the power which will be obtained in the definitive trial, define an appropriate test statistic, and use the statistic's sampling distribution to define and calculate type I and II error rates. We then use the method to re-design an external pilot trial comparing a psychotherapy with treatment as usual for children with conduct disorders in Section \ref{sec:example}. In Section \ref{sec:eval} we study the properties of the proposed test in a range of scenarios and compare its performance against the conventional approach to choosing pilot sample size and PCs. We extend the method in Section \ref{sec:extension} to allow for the additional goal of estimating the standard deviation of the primary outcome, before concluding with a discussion of implications and limitations in Section \ref{sec:discussion}.



\section{Problem and notation}\label{sec:problem}

Consider an external pilot planned in advance of a large two-arm parallel group trial which will compare an intervention with control based on a normally distributed primary endpoint. We assume that the definitive trial has a target sample size $n_t$ to be recruited from a pool of $n_e$ eligible patients. Each of the $n_e$ eligible patients will agree to take part in the trial with probability $\phi_r$, and recruitment will continue until either the target $n_t$ has been reached or the pool of eligible patients has been exhausted. We denote by $N$ the total number of participants in the definitive trial, a random variable with a truncated binomial distribution of size $n_e$, probability $\phi_r$, and constrained to be less than or equal to $n_t$.

We assume that the $N$ recruited participants of the definitive trial will be randomised equally between the two arms. In the control arm, $F_0$ participants will be successfully followed up with probability $\phi_f$. Thus, $F_0 ~|~ N \sim Bin(N/2, \phi_f)$. In the intervention arm, participants may or may not be followed-up, and may or may not adhere to the intervention. We allow for the possibility that these binary outcomes will be correlated at the participant level by using a multinomial distribution, such that participants will both adhere and be followed up with probability $p_{11}$; adhere, but be lost to follow up with probability $p_{10}$; not adhere, but be successfully followed up with probability $p_{01}$; and neither adhere nor be followed up with probability $p_{00}$. We can parametrise the model with marginal rate of follow-up $\phi_f$, rate of adherence conditional on being followed-up, $\phi_a$, and an odds ratio $\phi_{or}$:
\begin{align*}
p_{11} &= \phi_a \phi_f \\
p_{01} &= \phi_f - p_{11} \\
p_{00} &= \frac{\phi_{or} ~ p_{01}(1-p_{11}-p_{01})}{p_{11} + \phi_{or} ~ p_{01}} \\
p_{10} &= 1 - p_{11} - p_{01} - p_{00}.
\end{align*}
Note that we have assumed a constant marginal rate of follow-up in the intervention and control arms. We denote the number followed-up in the intervention arm by $F_1$, where $F_1 ~|~ N\sim Bin(N/2, \phi_f)$, and the number of those followed-up who also adhere by $A$, where $A ~|~ F_1 \sim Bin(F_1, \phi_a)$. 

We assume that non-adherence will be absolute in the sense that no treatment effect will be gained. We model the outcome for patient $i$ in arm $j$ as
$$
y_{i,j} = t_j a_{i,j} \mu + e_{i,j},
$$
where $t_i$ and $a_{i,j}$ are binary indicators of treatment arm and adherence respectively, $\mu$ is the difference in mean outcome between the two groups, $e_{i,j} \sim N(0, \sigma^2)$ is the residual term, and we have omitted the usual constant intercept for notational simplicity and without loss of generality. We will assume initially that $\sigma^2$ is known, although this will be relaxed in Section \ref{sec:extension}. For simplicity the primary analysis of the definitive trial is assumed to be a complete-case intention-to-treat $z$-test of the observed mean difference $\bar{Y}_1 - \bar{Y}_0$ with a null hypothesis of $H_0: \mu = 0$, where $\bar{Y}_j = \frac{1}{F_j} \sum_{i=1}^{F_j} y_{i,j}$.

We assume that the external pilot trial will take the same form as the definitive trial, but on a smaller scale. The model of recruitment, follow-up and adherence in the pilot trial is assumed to be the same as for the definitive trial apart from one aspect: for the small external pilot trial we will continue recruiting until the target pilot sample size of $n_p$ has been reached. Thus, we consider the achieved sample size of the pilot to be known and fixed, with the number of eligible patients approached but declining to participate following a negative binomial distribution $S \sim NB(n_p, \phi_r)$. Denoting the parameters by $\phi = (\phi_r, \phi_f, \phi_a)$, the external pilot trial will provide an estimate $\hat{\phi}$. Our goal is to show how a decision rule mapping $\hat{\phi}$ to the set $\{stop, go\}$ of progression decisions can be defined, and how the parameters of the rule and the pilot sample size $n_p$ can be chosen.

\section{A hypothesis test of feasibility}\label{sec:methods}

A decision rule mapping pilot estimates $\hat{\phi}$ to the set $\{stop, go\}$ of progression decisions can be specified and evaluated under a framework of hypothesis testing. Specifically, given a model of the pilot trial and thus of the sampling distribution for $\hat{\phi}$, we can calculate the probabilities of a given decision rule leading to `stop' and `go' decisions conditional on the true parameter values $\phi$. If we can identify those values $\phi$ for which we would like to make a `stop' decisions, and likewise those for which we would like to make a `go' decision, we can calculate long-run error rates of the first and second kind and use these to inform our choice of decision rule and of pilot trial sample size.

In this section we first propose that a planned definitive trial can be classified as either feasible or infeasible according to the power it will have in its final analysis. Power will be determined by the parameter $\phi$, and so this provides a means to define null and alternative hypotheses to be tested in the pilot trial. We begin by deriving an expression for the definitive trial power as a function of $\phi$, and then go on to define hypotheses. We then propose a test statistic to be used in the pilot trial and provide its sampling distribution, allowing long-run error rates to be calculated. Finally, we show how these error rates can be used to determine an optimal design for the pilot trial, encompassing both its sample size and the decision rule to be used at its conclusion.

\subsection{Power of the definitive trial}\label{sec:power}

The power of the definitive trial is determined by the sampling distribution of the difference in group means, $\bar{Y}_1 - \bar{Y}_0$. We assume that the definitive trial per-arm sample size will be greater than 30, allowing the sampling distributions of the group means to be approximated by normal distributions. The power of the definitive trial to detect a difference $\mu$ is then
$$
\Phi \left(\frac{\mathbb{E}[\bar{Y}_1 - \bar{Y}_0 ~|~ \mu]}{\sqrt{Var(\bar{Y}_1 - \bar{Y}_0 ~|~ \mu)}} - z_{1-\alpha} \right),
$$
where $\alpha$ denotes the (one-sided) type I error rate and $\Phi$ denotes the standard normal distribution function. The expectation and variance of the mean difference will depend on the values of $\mu, \phi, n_t$ and $n_e$. For clarity, we will drop the terms $\mu, n_t$ and $n_e$ from our notation as they can be considered fixed, and focus on power as a function of $\phi$, denoted $g(\phi)$. Then,
$$
g(\phi) = \Phi \left( x(\phi) - z_{1-\alpha} \right).
$$
The appendix will show that
$$
x(\phi) =  \frac{ \phi_a\mu \sqrt{\phi_f \mathbb{E}[N ~|~ \phi_r]} } {\sqrt{4\sigma^2 + 2 \mu^2 \phi_a(1-\phi_a)}},
$$
where $\mathbb{E}[N ~|~ \phi_r]$ is the expected number of participants recruited into the definitive trial when the recruitment rate is $\phi_r$,
$$
\mathbb{E}[N ~|~ \phi_r] = \sum_{k=0}^{n_t-1} k{n_e \choose k} \phi_r^k (1-\phi_r)^{n_e - k} + n_t Pr(C \geq n_t),
$$
and $C \sim Bin(n_e, \phi_r)$.

\subsection{Hypotheses and test statistic}

We propose to define null and alternative hypotheses by considering the power which would be obtained in the definitive trial which will aim to recruit $n_t$ participants from a pool of $n_e$ eligible patients. We require a power threshold $p_0$ to be identified, such that if the power of the definitive trial was known to be less than or equal to $p_0$ then it would be considered infeasible and we would like to minimise the chance of mistakenly proceeding to it (a type I error). Similarly, we require a second threshold $p_1$ such that if the definitive trial had a  power of at least $p_1$ we would considered it to be feasible and would like to minimise the chance of mistakenly concluding otherwise (a type II error). 

Given the thresholds $p_0, p_1$, we can define the null (alternative) hypothesis as those parameter values which would lead to a definitive trial with power less than or equal to $p_0$ (greater than or equal to $p_1$). We denote these hypotheses by $\Phi_0, \Phi_1$ respectively. We can express these in terms of the previously defined $x(\phi)$:
\begin{align*}
\Phi_0 &= \{\phi \in \Phi ~ | ~ x(\phi) \leq x_0 \} \\
\Phi_1 &= \{\phi \in \Phi ~ | ~ x(\phi) \geq x_1 \},
\end{align*}
where
$$
x_i = \Phi^{-1}(p_i) + z_{1-\alpha}.
$$

Testing feasibility in the external pilot trial will involve calculating a statistic based on the pilot estimate $\hat{\phi}$ and proceeding to the definitive trial if and only if it exceeds some pre-specified critical value $c$. A natural choice of statistic is $x(\hat{\phi})$. We can then define the type I and II error rates of the pilot trial, which will be determined by the pilot sample size $n_p$ and the critical value $c$:
\begin{align*}
\alpha(n_p, c) = \max_{\phi \in \Phi_0} Pr[ x(\hat{\phi}) > c ~ | ~ \phi, n_p], \\
\beta(n_p, c) = \max_{\phi \in \Phi_1} Pr[ x(\hat{\phi}) < c ~ | ~ \phi, n_p].
\end{align*}

\subsection{Power of the external pilot trial}

To obtain type I and II error rates for a particular pilot trial of sample size $n_p$ per arm and critical value $c$, we require an expression for the probability that the pilot test statistic will be greater than $c$ conditional on some true parameter value $\phi$, denoted
$$
h(n_p, c, \phi) \equiv Pr[x(\hat{\phi}) > c ~|~ n_p, \phi].
$$
Denote by $n_{af}$ the value taken of the multinomial variable recording follow-up and adherence outcomes, and by $\mathcal{N}_{af}$ the set of possible realisations. The power of the pilot trial can be calculated by considering each value in $\mathcal{N}_{af}$ and calculating the probability that the observed recruitment rate will be sufficient for the test $x(\hat{\phi}) > c$ to pass. Given adherence and follow-up estimates $\hat{\phi}_a, \hat{\phi}_f$, this condition will be met when the estimated recruitment rate $\hat{\phi}_r$ is such that
$$
E[N ~|~ \hat{\phi}_r] > \frac{c^2(4 + 2 \mu^2 \hat{\phi}_a (1- \hat{\phi}_a)}{\mu^2 \hat{\phi}_a^2 \hat{\phi}_f} = \tilde{n}.
$$ 
Since $\hat{\phi}_r = 2n_p/(2n_p + s)$, we can find the largest value of $s$ such that $E[N ~|~ \hat{\phi}_r] > \tilde{n}$. Noting this will be a function of $n_{af}$ and denoting it by $\tilde{s}(n_{af})$, the power of the pilot trial is then
\begin{align}\label{eqn:pilot_pow}
h(n_p, c, \phi) &= \sum_{n_{af} \in \mathcal{N}_{af}} 
Pr\left[S \leq \tilde{s}(n_{af}) ~|~ \phi \right] 
p_{af}(n_{af} ~|~ \phi).
\end{align}
Here, $p_{af}(.)$ is the multinomial density describing follow-up and adherence outcomes.

Equation \ref{eqn:pilot_pow} allows for follow-up and adherence outcomes to be correlated, which may be appropriate if we expect that a participant who does not adhere to the intervention delivery is also less likely to adhere to other aspects of the trial protocol, including data collection. If this is not thought to be the case, we can assume follow-up and adherence are independent and equation \ref{eqn:pilot_pow} simplifies to
\begin{equation*}
h(n_p, c, \phi) = \sum_{a=0}^{n_p} \left(  \sum_{f=0}^{2n_p} 
Pr\left[S \leq \tilde{s}(a,f) ~|~ \phi \right] 
p_{f}(f ~|~ \phi)  \right) p_a(a ~|~ \phi),
\end{equation*}
where $p_f(.)$ and $p_a(.)$ are the densities describing the follow-up and adherence outcomes respectively. By avoiding a summation over the multinomial space $\mathcal{N}_{af}$, which can be very large for moderate $n_p$, this reduces the number of terms to be evaluated and thus decreases computation time.


\subsection{Operating characteristics}\label{sec:ocs}

Recall that the type I (II) error rate of the pilot trial is defined as the largest probability of proceeding (failing to proceed) to the definitive trial when the true parameter is in the null (alternative) hypothesis. That is, 
\begin{align*}
\alpha(n_p, c) &= \max_{\phi \in \Phi_0} ~ h(n_p, c, \phi) , \\
\beta(n_p, c) &= \max_{\phi \in \Phi_1}  ~ 1 - h(n_p, c, \phi).
\end{align*}

Designing a pilot trial could proceed by considering a set of candidate sample size values $n_p$ and critical values, solving the above optimisation problems for each, and then choosing the $(n_p, c)$ pair which is deemed to give the best balance between the costs of sampling and the two types of errors. This approach has the drawbacks of requiring an appropriate range of $c$ \emph{a priori}, and a lack of sharing information between the discrete optimisation problems. An alternative formulation which avoids these drawbacks is to (for fixed $n_p$) cast the problem as one of constrained bi-objective optimisation over the space $\mathbb{R} \times \Phi \times \Phi$, simultaneously searching for a critical value $c$ and two points in the parameters space, $\phi_0$ and $\phi_1$, which maximise type I and II error rates whilst satisfying the constraints that $\phi_0 \in \Phi_0$ and $\phi_1 \in \Phi_1$. Formally, we solve

\begin{alignat}{1}\label{eqn:MO_opt}
\max ~ & \left( h(n_p, c, \phi_0) , ~ 1 - h(n_p, c, \phi_1)  \right) \\
\text{subject to} ~ & c \in \mathbb{R}, \nonumber \\ 
& \phi_0 \in \Phi_0, \nonumber  \\
& \phi_1 \in \Phi_1. \nonumber 
\end{alignat}

A problem of this nature can be solved numerically using the NSGA-II algorithm \cite{Deb2002}, as implemented in the R package `mco' \cite{Mersmann2014}. It will provide a set of critical values and corresponding points in the null and alternative hypotheses offering different balances between type I and type II error rates. These error rates can then be plotted for a number of choices of $n_p$, and an appropriate design selected from them. To ensure a reasonably fast computation time, we implemented the function $h(n_p, c, \phi)$ in C++ via the `Rcpp' package\cite{Eddelbuettel2011}. Full details, including all code used to produce the results described throughout the paper, are provided in the supplementary materials.



\subsection{Comparison}\label{sec:comparator}

An alternative to the proposed test is to follow the conventional approach and make the stop/go decision based on several independent progression criteria. Decision rules are then defined by three critical values $\mathbf{c} = (c_f, c_a, c_r)$, where we proceed to the definitive trial only when $\hat{\phi}_f > c_f$, $\hat{\phi}_a > c_a$ and $\hat{\phi}_r > c_r$. Assuming independence between the parameter estimates, the probability of this event is
\begin{align*}
f(n_p, \mathbf{c}, \phi) &= Pr[\hat{\phi}_r > c_r ~ | ~ \phi] \times Pr[ \hat{\phi}_f > c_f ~ | ~ \phi] \times Pr[ \hat{\phi}_a > c_a ~ | ~ \phi] \\
&= F_s( 2n_p/c_r - 2n_p ~ | ~ \phi) \times [1-F_f(2n_p c_f ~ | ~ \phi)] \times [1-F_a(n_p c_a ~ | ~ \phi)],
\end{align*}
where $F_s(.), F_f(.)$ and $F_a(.)$ are the cumulative distribution functions for the random variables $S, F$ and $A$ in the pilot, respectively. For any given choice of pilot sample size $n_p$ and progression criteria $\mathbf{c}$, type I (II) error rates are defined as before by maximising $f(n_p, \mathbf{c}, \phi)$ over the null (alternative) hypotheses. To find a set of progression criteria which offer different trade-offs between minimising type I and II error rates, we solve the following bi-objective optimisation problem:

\begin{alignat*}{1}
\min_{\mathbf{c} \in [0,1]^3} ~ & \left( \max_{\phi \in \Phi_0} f(n_p, \mathbf{c}, \phi) , ~ \min_{\phi \in \Phi_1} 1 - f(n_p, \mathbf{c}, \phi) \right). \\
\end{alignat*}

We use NSGA-II again to solve the outer bi-objective minimisation problem. For the inner optimisation problems, we assume that the solutions will lie on the boundaries of the hypotheses and therefore have a constraint, $\phi \in \Phi_i$, meaning that the search is over two dimensions. For example, we can search over $(\phi_r, \phi_a) \in [0,1]^2$ since $\phi_f$ will then be determined by
$$
\phi_f = \frac{4\sigma^2 + 2\mu^2 \phi_a(1-\phi_a)x_i^2}{(\phi_a \mu)^2 E[N ~|~ \phi_r]}.
$$
Given the low dimension of the search space, we obtain approximate solutions by searching over a fine grid of values for $(\phi_r, \phi_a)$, in increments of 0.005. This procedure is computationally efficient since the error rates at each point in the grid can be calculated in a vectorised manner, with very little overheads in comparison to a sequential evaluation. Again, full details are provided in the supplementary materials. 

\section{Application to an external pilot trial of a psychotherapy intervention}\label{sec:example}

TIGA-CUB \cite{Edginton2017} was a two-arm, parallel group, individually-randomised external pilot trial aiming to determine the feasibility of a definitive trial comparing second-line, short-term, manualised psychoanalytic child psychotherapy with treatment as usual for children with conduct disorders. The trial's objectives included estimation of the rate at which eligible dyads (where a child - carer dyad was the unit of randomisation and observation) consented to take part in the trial; the level of missing data in the primary outcome; the rate of adherence to the intervention; and the parameters required for the sample size calculation of the main study. Progression criteria relating to recruitment, follow-up and adherence rate were specified, with the definitive trial to go ahead only if:
\begin{enumerate}
\item Recruitment was to target;
\item Attendance was at more than 50\% of sessions in the intervention arm;
\item At least 75\% of follow-up data was collected.
\end{enumerate}

The sample size was determined by using the rule-of-thumb that 30 participants per arm is sufficient to estimate a standard deviation of a continuous primary outcome which is common across arms \cite{Lancaster2004}. Assuming 90\% of pilot participants are followed-up, a sample size of 60 participants in total was chosen and justified in terms of the expected width of 95\% confidence intervals around estimates of the standard deviation (0.39 multiplied by SD), the follow-up rate ($\pm 7$\%), and the adherence rate (between $\pm 8$\% to 18\%).

To revisit the choice of sample size and progression criteria, we first need to define some aspects of the planned definitive trial. We begin by setting the minimal clinically important effect size to be detected and the outcome standard deviation, here assumed to be $\mu = 0.3$ and $\sigma^2 = 1$ respectively. We then decide on the planned design of the definitive trial, in terms of the number of eligible patients who will be approached and the target total sample size. For illustrative purposes we suppose these are equal to $n_e = 1000$ and $n_t = 514$. Note that this target sample size, after allowing for 10\% attrition, would lead to a definitive trial with 90\% power to detect $\mu=0.3$ when using a two-sided type I error rate of 0.05.

\begin{figure}
\centering
\includegraphics[scale=0.8]{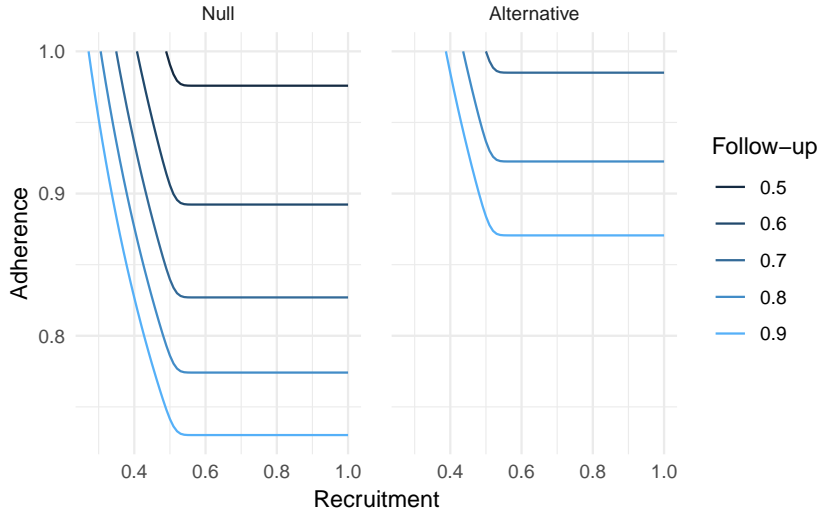}
\caption{Values of recruitment, follow-up and adherence parameters for leading to a definitive trial power of $p_0 = 0.65$ (i.e. the boundary of the null hypothesis, left panel), or $p_1 = 0.8$ (i.e. the boundary of the alternative hypothesis, right panel).}
\label{fig:hyps}
\end{figure}

We decide that if the planned definitive trial were to have at least 80\% power, after accounting for any issues with recruitment, follow-up or adherence, we would like to obtain a `go' decision from the pilot trial. In terms of our notation, we set $p_1 = 0.8$. Similarly, we consider that if the trial could only provide a power of 65\% or less then we would like to obtain a `stop' decision from the pilot, and so we set $p_0 = 0.65$. These parameters specify the null and alternative hypotheses which are to be tested in the pilot trial. They can be visualised by considering pairs of recruitment and adherence rates and finding the corresponding value of follow-up rate that would lead to a definitive trial power of exactly $p_0 = 0.65$ for the null, or $p_1 = 0.8$ for the alternative. The resulting surfaces, which represent the boundaries of the hypotheses, can then be plotted as in Figure \ref{fig:hyps}. In this case, we see that the two hypotheses have a similar shape, but with the alternative shifted towards higher (and therefore better) values of follow-up, recruitment and adherence rates. The trade-offs between the three parameters are clear, with decreases in one being compensated by increases in the others in order to maintain power at either $p_0$ or $p_1$. As the recruitment rate increases beyond around 50\% there are no opportunities for trade-offs with the remaining parameters because it becomes certain that the recruitment target of $n_t$ will be met, and so the sample size will not increase any further.

\subsection{Fixing $p_0$ and $p_1$}

\begin{figure}
\centering
\includegraphics[scale=0.8]{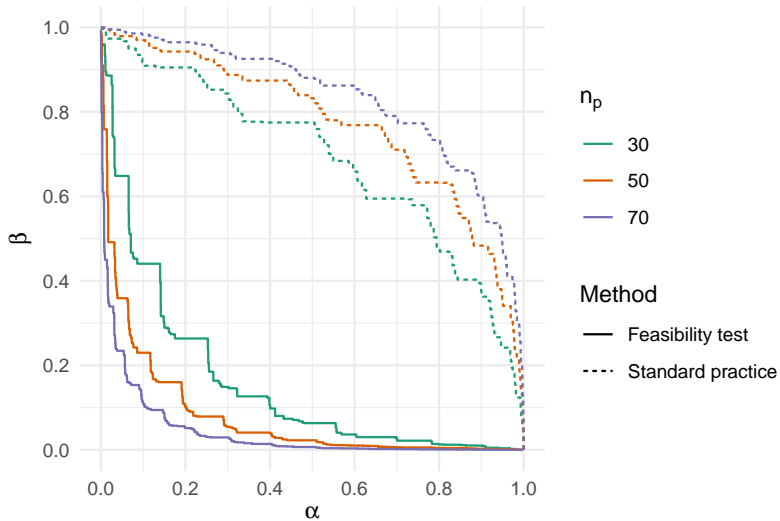}
\caption{Type I ($\alpha$) and type II ($\beta$) error rates obtained for a range of critical values $c$ and pilot sample sizes $n_p$ when using the proposed method (solid lines). Error rates available when using conventional progression criteria are shown for comparison (dashed lines).}
\label{fig:ex_ocs}
\end{figure}

Given the hypotheses, and assuming that follow-up and adherence rates are independent, we can now consider different sample sizes and progression criteria and calculate the error rates they lead to. For illustration, we consider pilot trial sample sizes of $n_p = 30, 50, 70$ per arm and solve the optimisation problem \ref{eqn:MO_opt} for each case. The resulting operating characteristics are plotted in Figure \ref{fig:ex_ocs} (solid lines). In this case we see that error rates with the original sample size of $n_p = 30$ are quite poor, in comparison to the nominal values typically seen in early phase drug trials. For example, a type I error rate of 0.09 corresponds to a type II error rate of 0.44. As we would expect, increasing the external pilot sample size leads to improved error rates. A pilot sample size of 50 participants per arm, for example, reduces the type II error rate to around 0.23 whilst maintaining type I error rate at 0.09. We might conclude from these results that TIGA-CUB's original sample size of 30 per arm was too small to make reliable progression decisions, and that increasing to somewhere between 50 and 70 per arm could have been worthwhile.

Given the possible choices of error rates and sample size given in Figure \ref{fig:ex_ocs} we can select that which best reflects our priorities. For example, suppose we select the point with $n_p = 50, \alpha = 0.09$ and $\beta = 0.23$. The corresponding critical value is $c = 2.6422$, so the pilot trial will produce a `go' decision if $x(\hat{\phi}) > 2.6422$ and a `stop' decision otherwise. It may be helpful to transform the decision rule to the power scale, allowing us to specify the decision rule as: `go' if the predicted power of the main trial, based on the pilot estimate $\hat{\phi}$, is greater than $\Phi(2.6422 - z_{1-0.025}) = 0.7524$; otherwise, `stop'.

For comparison, we also evaluated the error rates resulting from the conventional approach to defining PCs described in Section \ref{sec:comparator}. For each of the pilot sample sizes $n_p = 30, 50, 70$ we searched over the full set of possible PC thresholds $\mathbf{c}$ to identify a non-dominated subset, where a threshold $\mathbf{c}$ is non-dominated if there does not exist another threshold $\mathbf{c}'$ which leads to lower type I \emph{and} lower type II error rates. In this example we find that setting several independent PCs will lead to very poor error rates. For example, when $n_p = 30$  we must accept a type II error rate of 0.9 (i.e. a power of 0.1) to obtain a type I error rate of 0.1. In fact, regardless of the specific thresholds used, error rates will be no better than what would be obtained when making progression decisions by tossing a (possibly biased) coin. Increasing the pilot sample size does not improve error rates, but actually makes them worse. These counter-intuitive results can be explained by the fact that independent PCs are of an intrinsically \emph{conjunctive} nature, requiring all three endpoints to show a positive result. In contrast, our hypotheses have been defined in a more \emph{disjunctive} manner, where a positive result in only one endpoint may be enough to indicate overall feasibility due to the acknowledged trade-offs involved. To illustrate the point further, consider the case of a  pilot trial sample size tending to infinity so that $\hat{\phi} = \phi$ and pilot error rates will be either 0 or 1. In this idealised scenario, a type II error rate of $\beta = 0$ will be obtained only if $c_f < \phi_f, c_a < \phi_a, c_r < \phi_r ~ \forall ~ \phi \in \Phi_1$. For this infinite sample size to also lead to $\alpha = 0$, a necessary condition is therefore that
$$
\phi' = \left( \min_{\Phi_1} ~ \phi_f, ~ \min_{\Phi_1} ~ \phi_a, ~ \min_{\Phi_1} ~ \phi_r \right) \not\in \Phi_0.
$$ 
For example, in this case we find $\phi' = (\phi_f'=0.679, \phi_a'=0.83, \phi_r'=0.35)$, but $x(\phi') = 1.91 < x_0$ and so $\phi'$ \emph{is} in the null hypothesis. Thus, a pilot sample size tending to infinity and with independent PC thresholds chosen to give $\beta = 0$ will have $\alpha = 1$.

Aside from the overall inefficiency of independent PCs, we found that there are many possible choices for independent PC thresholds which are dominated by others. For example, for $n_p = 30$ we find that $\mathbf{c} = (c_f=0.705, c_a=0.865, c_r=0.373)$ leads to type I and II error rates of 0.53 and 0.72 respectively, whereas $\mathbf{c}' = (c_f=0.6, c_a=0.8, c_r=0.4)$ will give a worse type I error (0.74) \emph{and} a worse type II error (0.88). This suggests that if independent PCs are to be used under the conventional approach, as they were in TIGA-CUB, the choice of threshold values should be informed by a careful statistical analysis of their properties if we are to avoid needlessly inefficient decision rules. 

\subsection{Fixing $p_1$ and $\beta$}

An alternative approach to pilot trial design results from the observation that, if the alternative hypothesis threshold $p_1$ is fixed, we can find a critical value $c$ which will lead to some desired pilot type II error rate independently of $p_0$. Keeping $c$ fixed at this value, we can then look at a range of values for $p_0$ and plot the corresponding type I error rate. The evaluation of a pilot with sample size $n_p$ and critical value $c$ can then be based on the probability of making a go decision as a function of $p_0$, rather than focussing on a specific point.

\begin{figure}
\centering
\includegraphics[scale=0.8]{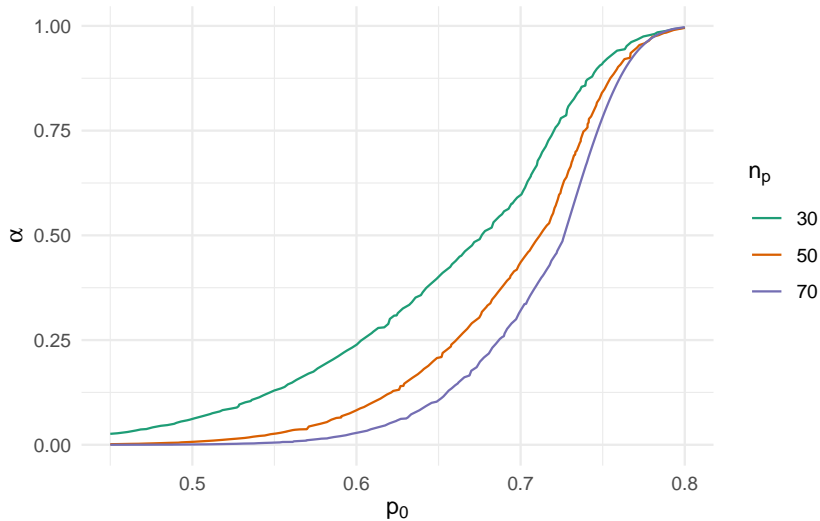}
\caption{Type I error rates for pilot trials of different sample sizes $n_p$, as a function of the null hypothesis parameter $p_0$ and keeping type II error maintained at 0.1.}
\label{fig:alt}
\end{figure}

For example, suppose we take $p_1 = 0.8$ as before and ask for a type II error rate of 0.1. We find that, for a pilot sample size of 30 per arm, a critical value of  $c = 2.46$ will achieve this error rate. We can then keep $c$ fixed at this value and calculate the corresponding type I error rate as we vary $p_0$. Doing this again for pilot sample sizes of 50 and 70, we plot the results in Figure \ref{fig:alt}. We see that while a large pilot sample of $n_p = 70$ will ensure there is only a 0.03 probability of proceeding to a definitive trial with a true power of $p_0 = 0.6$, it will also mean that when $p_0 \approx 0.725$ there will only be a 0.5 chance of proceeding. This might be considered rather low, when we have defined the alternative hypothesis using $p_1 = 0.8$. In comparison, the associated values when $n_p = 30$ are 0.24 and 0.78.

\section{Evaluation}\label{sec:eval}

To understand to what extent the results given in the above illustrative example will be found more widely, we applied both the proposed test and the conventional method to a number of different scenarios. Recall that a scenario is defined by: the parameters describing the primary outcome of the definitive trial (mean $\mu$ and variance $\sigma^2$); its planned design (target sample size $n_t$ and number of eligible patients $n_e$); and the power thresholds used to define our null and alternative hypotheses ($p_0$ and $p_1$ respectively). Scenarios are distinguished from one another in terms of the null and alternative hypotheses they imply, which are determined by $x(\phi)$ and the thresholds $p_0, p_1$. 

\begin{figure}
\centering
\includegraphics[scale=0.8, trim={1.5cm 0 0 0},clip]{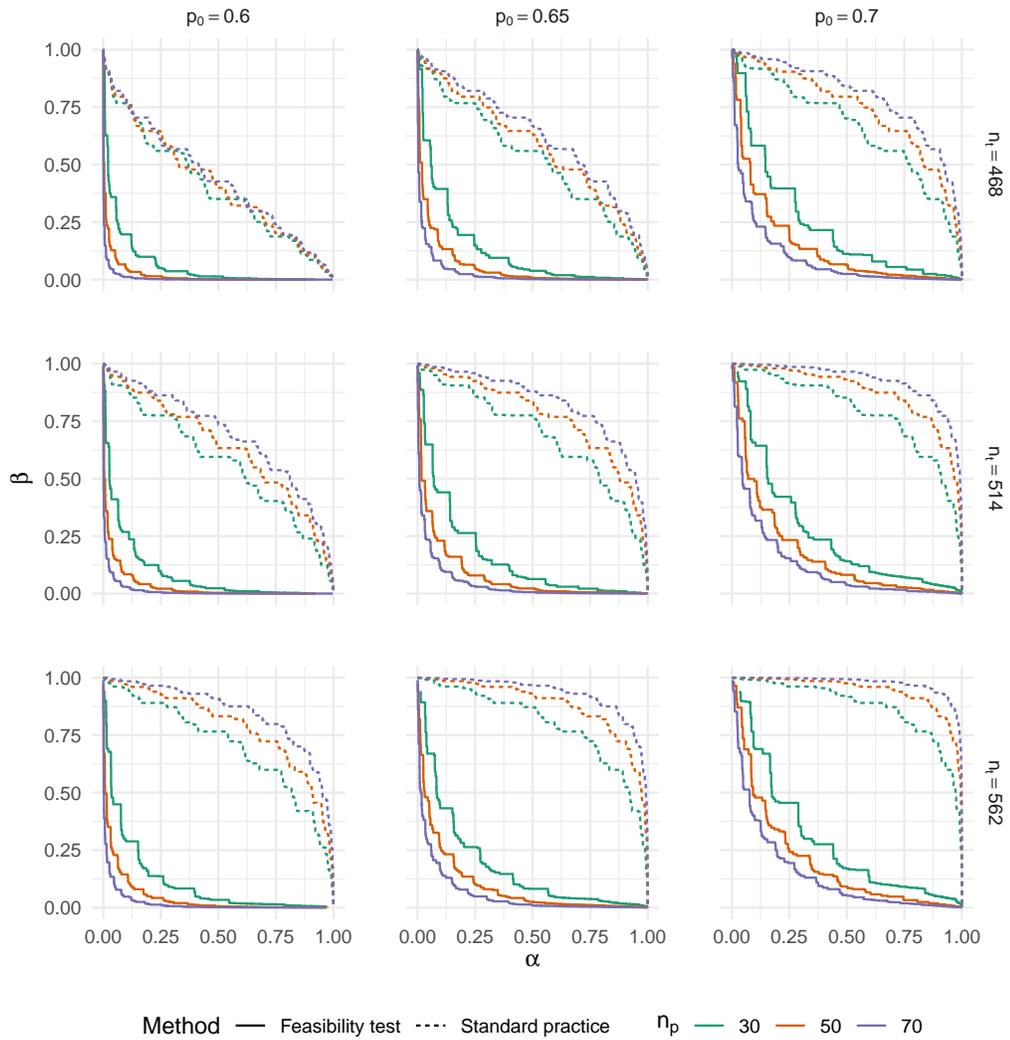}
\caption{Type I ($\alpha$) and type II ($\beta$) error rates obtained for a range of critical values $c$ and pilot sample sizes $n_p$ when using the proposed method (solid lines) and the conventional approach (dashed line). The power used to define the null hypothesis increases from left to right, $p_0 = 0.6, 0,65, 0.7$, while the target sample size of the definitive trial is increased from top to bottom, $n_t = 468, 514, 562$.}
\label{fig:eval}
\end{figure}

We will fix $p_1 = 0.8$ throughout, and will consider $p_0 = 0.6, 0.65, 0.7$. We note that the denominator in $x(\phi)$ will be dominated by its first term for the typical standardised effect sizes (i.e less than around 0.5). The expected definitive trial sample size in the numerator depends on $n_e, n_t$ and $\phi_r$ and so, for fixed $\phi$, the terms $\mu, \sigma^2, n_e$ and $n_t$ reduce to one factor when determining $x(\phi)$. As such, we focus on varying one of these parameters while keeping the others fixed. We choose to fix $\mu = 0.3, \sigma^2 = 1, n_e = 1000$ whilst varying $n_t$. We consider $n_t =  468, 514, 562$, corresponding to the target sample size which would be obtained when asking for 90\% power and allowing for attrition of 0, 10 and 20\% respectively. For each scenario we calculated the operating characteristics of an external pilot with sample size $n_p = 30, 50, 70$ per arm. The resulting error rates are illustrated in Figure \ref{fig:eval}.

We see that inflating the target sample size leads to a small increase in pilot error rates. For example, for $n_p = 30$, $n_t = 468$ and $p_0 = 0.7$ (top right panel), the pilot can have error rates of $\alpha = 0.08$ and $\beta = 0.58$. Increasing the target sample size to $n_t = 562$ gives an increased type II error rate of $\beta = 0.77$ whilst maintaining type I error. This general behaviour can be explained by noting that recruiting more participants will mean we can tolerate worse follow-up and/or adherence whilst maintaining power. As a result the hypotheses will be larger, and in general we can expect error rates to increase as the size of the hypotheses they are defined over increase. Error rates are also sensitive to the power used to define the null hypothesis. For example, with pilot sample size $n_p = 30$ and definitive trial target sample size $n_t = 468$ but reducing $p_0$ from 0.7 to 0.6, one can reduce the type II error rate from 0.58 to 0.20 at the cost of a small increase in the type I error rate from 0.08 to 0.11. In all scenarios we see a considerable improvement in error rates when we increase the pilot sample size from 30 to 50 per arm, with less of a subsequent improvement moving from 50 to 70. Although the choice of sample size for a given pilot trial will depend on the costs of sampling the consequences of type I and II errors, our results suggest that a sample size of around 50 per arm should be sufficient whenever $p_1 = 0.8$ and $p_0 \leq 0.65$.

With regards to the conventional approach of setting several independent PCs, we see similar trends in terms of the effects of changing $n_t$ and $p_0$. As with the proposed method, error rates tend to increase as $n_t$ increases, but in this case at a higher rate. Reducing $p_0$ leads to improved error rates. Even in the best scenario we considered, with $n_t = 468$ and $p_0 = 0.6$, the error rates which can be obtained using independent PCs are still no better than those of a random coin toss. We can conclude from these results that, when the proposed method of defining hypotheses is deemed appropriate, independent PCs should not be used to guide progression decisions in external pilot trials.

\section{Extension - unknown variance}\label{sec:extension}

We have considered how recruitment, follow-up and adherence rates all affect the power of the definitive trial, and have shown how these can be assessed in an external pilot trial as part of a hypothesis test. Another parameter which affects the power of the definitive trial is the standard deviation of the outcome measure: recall from Section \ref{sec:power} that the conditional power is 
$$
g(\phi, \sigma) = \Phi \left( x(\phi, \sigma) - z_{1-\alpha} \right),
$$
where
$$
x(\phi, \sigma) =  \frac{ \phi_a\mu \sqrt{\phi_f \mathbb{E}[N ~|~ \phi_r]} } {\sqrt{4\sigma^2 + 2 \mu^2 \phi_a(1-\phi_a)}},
$$
The methods described so far have assumed $\sigma$ is known. When this is not the case we can extend our approach by including the true value of $\sigma$ in the definition of the hypotheses $\Phi_0$ and $\Phi_1$, and the pilot estimate $\hat{\sigma}$ in the test statistic $x(\hat{\phi}, \hat{\sigma})$. This will allow for high variability in the outcome measure, which may lead to a trial with infeasibly low power, to be identified at the pilot stage.

For notational simplicity and ease of computation we will focus on the case where follow-up and adherence are independent, but the method extends naturally to the more general case. The power of the pilot trial is now approximately 
\begin{equation*}
h(n_p, c, \phi, \sigma) = \sum_{a=0}^{n_p} \left[  \sum_{f=0}^{2n_p} \left( \sum_{s = 0}^{s_{max}} 
\left[ \int_0^y p_{\hat{\sigma}^2}( \hat{\sigma}^2 | s, a, f, \phi) ~ d\hat{\sigma}^2 \right] 
p_s(s | \phi) \right) p_{f}(f |\phi)  \right] p_a(a | \phi),
\end{equation*}
where $p_{\sigma^2}(.)$ is the density function of the sample variance, specifically,
$$
\hat{\sigma}^2 ~|~  f \sim \frac{\sigma^2 \chi^2_{f - 1}}{f - 1}.
$$
The approximation comes from the fact that we would like to sum over the full range of possible values for $s \in [0, \infty)$ but must limit the summation to some value $s_{max}$. We set $s_{max}$ to the upper 0.999 quartile of the negative binomial distribution with parameter $\phi_r$, ensuring an accurate approximation whilst avoiding excessive computation. The upper limit of the integral, $y$, is the largest value that the sample variance could take and still lead to the test $x(\hat{\phi}, \hat{\sigma}^2) > c$ passing, given $s, a$ and $f$. It is:
$$
y = \frac{\hat{\phi}_a^2 \mu^2 \hat{\phi}_f \mathbb{E}[N ~|~ \hat{\phi_r}]}{4c^2} - \frac{1}{2} \mu^2 \hat{\phi}_a (1-\hat{\phi}_a).
$$
 
Empirically, we find that both type I and II error rates increase as the standard deviation decreases. To avoid the trivial situation where error rates tend to 1 as $\sigma$ tends to zero, we include a lower limit on $\sigma_*$ in the definitions of the hypotheses $\Phi_0$ and $\Phi_1$. These are now defined as
\begin{align*}
\Phi_0 &= \{\phi \in \Phi, \sigma > \sigma_*  ~ | ~ g(\phi, \sigma) <= p_0 \}, \\
\Phi_1 &= \{\phi \in \Phi, \sigma > \sigma_* ~ | ~ g(\phi, \sigma) >= p_1 \}.
\end{align*}

To illustrate the effect of allowing for uncertainty in $\sigma$, we calculated the error rates available when $n_p = 50, n_t = 514, p_0 = 0.65$ and $p_1 = 0.8$, and the lower limit was taken to be $\sigma_* = 0.8$. We then compared these error rates with those previously obtained when it was assumed that $\sigma = 1$. For example,when $\sigma$ was assumed known a pilot trial type I error rate of 0.12 allowed a type II error rate of 0.23. Maintaining the same type I error but now estimating $\sigma$ in the pilot increases type II error to 0.39. The detriment stems from two issues. Firstly, by using the pilot variance estimate in the test statistic $x(\hat{\phi}, \hat{\sigma})$ its sampling variability increases. Secondly, by allowing for the lower limit of $\sigma_* = 0.8$ we accommodate lower values of the rates $\phi_r, \phi_a$ and $\phi_f$ in our hypotheses, enlarging them and thus allowing more extreme error rates to be located. Error rates can be improved to some extent through increasing sample size. For example, increasing $n_p$ from 50 to 70 will lead to a type II error rate of 0.34 whilst maintaining the type I error rate at 0.12.

\section{Discussion}\label{sec:discussion}

We have proposed a statistical test which simultaneously assesses recruitment, follow-up and adherences rates in order to anticipate and mitigate related problems which would render the planned definitive trial infeasible. We have shown that, without increasing the sample size of the pilot trial beyond typical values, the test can limit the probability of mistakenly running an underpowered trial whilst reliably ensuring well-powered trials will progress. We have described how the test can be extended to include the common pilot objective of estimating an unknown standard deviation nuisance parameter, and found that for this to be incorporated the sample size of the pilot trial will have to be increased if operating characteristics are to be maintained. 

The proposed method leads to reasonable error rates when the pilot sample size is around 50 patients per arm. While this is more than some common rules-of-thumb for choosing an external pilot sample size such as 25 \cite{Whitehead2015} or 35 \cite{Teare2014} per arm, it remains reasonable in the context of a pilot preceding a much larger trial. In comparison, our study of the conventional approach to setting progression criteria has shown that, at least in the scenarios we considered, no amount of increase to the pilot sample size will lead to reasonable error rates. Although the proposed method is more complicated than the conventional approach, it is flexible in the sense that it can be applied in its full generality when the aim of the pilot is to estimate recruitment, adherence, follow-up and the outcome standard deviation, but could equally be used in the special case where only a subset of these parameters are of interest. We have produced a robust and efficient implementation in R and have provided this in the supplementary material, where we also show how to reproduce the results given in this paper.

Some assumptions have been made for simplicity and ease of explanation. In particular, we assumed that the rate of follow-up will be constant across the intervention and control arms of the trials. Relaxing this assumption is possible, although it would add another dimension when searching for the maximum error rates of a given pilot sample size and critical value. We also assumed that adherence is absolute, with non-adherers receiving no treatment effect. Violation of this assumption will lead to a general under-estimation of definitive trial power, which will affect the construction of the pilot hypotheses and subsequently the pilot error rates. Finally, we note assumptions of a complete-case ITT analysis (as we have done for the definitive trial) and of the successful recruitment of the target sample size (as we have done for the pilot trial) are both standard when carrying out power calculations.

We have focussed on a specific problem where the definitive trial will have a randomised two-arm design with a single, normally distributed primary outcome measure. Our method could be applied directly to any problem with an outcome that can be approximately modelled as normal. For example, a binary endpoint can be incorporated using a normal approximation if the definitive trial sample size is sufficiently large; or a cluster randomised trial could be addressed if the endpoints are all measured at the cluster level and the variance parameters of the primary outcome were known. To extend the method to other problems we would require an expression for the power of the definitive trial (as a function of parameters $\phi$), a statistic to use in the pilot trial, and an expression for the power of the pilot trial using that statistic. When analytic forms of the power functions cannot be found, they can be approximated using simulation \cite{Landau2013} and, in principle, the optimisation problem in Section \ref{sec:ocs} could be solved as before. In practice, evaluating two objectives and two constraints using simulation would be computationally demanding and the NSGA-II algorithm would take an infeasibly long time to converge. Future work could explore how efficient global optimisation algorithms \cite{Jones2001} could be used to solve these problems in a timely manner. A further extension could consider internal pilot trials, where the pilot data is combined with the definitive data in the final analysis. This would, however, induce a correlation between the pilot trial progression decision and the final test of treatment effectiveness. As a result, appropriate adjustments would be needed to ensure the type I error rate of the final analysis can be controlled at the nominal level.

Our method requires the specification of the number of eligible patients $n_e$ and the target sample size $n_t$ for the definitive trial. In practice, the exact value of these parameters may not be known at the pilot design stage, and in particular the results of the pilot may influence how they are set. However, note that $n_e$ and $n_t$ are only used, in conjunction with the power thresholds $p_0, p_1$, to specify the null and alternative hypotheses $\Phi_0, \Phi_1$. Thus, we are free to use some hypothetical choices of $n_e$ and $n_t$ providing we accept that any point $\phi \in \Phi_i, i=0,1$ will remain as such even if $n_e$ or $n_t$ are changed. For example, we may have defined the null hypothesis so that $\phi \in \Phi_0$ if and only if a definitive trial with design $n_t, n_e$ would have less than 60\% power under $\phi$. For that same $\phi$, we could increase the definitive trial sample size to some $n_e', n_t'$ such that its power would increase to 90\%. Under our formulation we would still consider $\phi$ to be in the null, and would want to avoid running a definitive trial with sample size $n_e', n_t'$ and power $g(\phi, n_e', n_t') = 0.9$. This corresponds to a judgement that the improvement in power obtained by moving from $n_e, n_t$ to $n_e', n_t'$ is not sufficient to justify the increased cost of sampling. 

Our method leads to a binary stop/go decision. Increasingly, progression criteria in pilot trials are incorporating three outcomes, adding an additional intermediate `amber' decision between the `red' stop and `green' go \cite{Avery2017}. The intention is that if the pilot estimate is neither clearly good nor bad, but somewhere between, then it may be possible to make some modifications to the intervention or the trial protocol which would ensure the definitive trial will be feasible. A possible extension to the testing framework outlined could consider defining three hypotheses where the decisions `stop', `modify' or `go' would be optimal. In practice, the pilot trial could be designed using the proposed method and assuming no modifications will be possible, and, in the event that we wish to make modifications, another pilot could be done to assess the effect of these. This would appear to be in line with the iterative process of complex intervention development and evaluation suggested in the MRC framework\cite{Craig2008}. Alternatively, if the modifications are made and we proceed directly to the definitive trial, we should be clear that the error rates associated with the pilot trial as designed will no longer apply.

In addition to the recruitment, follow-up and adherence rates and the standard deviation we have considered in this paper, another parameter which would be of interest when deciding to run a definitive trial is the treatment effect $\mu$ \cite{Wilson2015}. In particular, if the true treatment effect is very small, we would ideally like to recognise this at the pilot stage and declare futility. To incorporate such an assessment into our method we could consider a bivariate test in the pilot, simultaneously assessing the power of the definitive trial and the true treatment effect. Defining the null and alternative hypotheses for such a test may be challenging, as it would require specifying the trade-offs we would be willing to make between definitive trial power and treatment effect. Future research could consider how methods for testing two outcomes in phase II drug trials\cite{Conaway1996, Thall2008} could be applied in this context.

The proposed method could benefit from further methodological research. For example, the pilot test statistic was defined in a pragmatic manner and is not guaranteed to be optimal, and so further research could consider alternative statistics which might be more powerful when testing the hypotheses we have defined. A further criticism of the proposed method is the simplicity of the decision rule it suggests. Particularly in the case of external pilot trials of complex interventions, we would expect the decision of if and how the definitive trial should be conducted to be informed by many factors beyond the pilot estimates of a handful of parameters, not least qualitative outcomes. However, we believe the method will still provide a useful guide to decision making, and allows for the choice of pilot sample size to be considered more fully. This view is supported by the continued prevalence of hypothesis testing in the design and analysis of all types of clinical trials, where the final decision is rarely made by rigidly following the result of the test. 

\subsection*{Acknowledgements}

We would like to thank Alex Wright-Hughes and the TIGA-CUB trial team for discussions which helped shape the scope of this paper.

\subsection*{Data availability statement}

Data sharing is not applicable to this article as no new data were created or analysed in this study. The code used to implement the methods and generate the results of this paper is freely available at 

\subsection*{Funding}

This work was supported by the Medical Research Council [grant number MR/N015444/1].

\bibliographystyle{unsrt}
\bibliography{C:/Users/meddwilb/Documents/Literature/Databases/DTWrefs}

\section*{Appendix}

Recall that the sample means are
\begin{align*}
\bar{Y}_1 &= \frac{A}{F_1}\mu + \frac{1}{F_1} \sum_{i=1}^{F_1} e_{i,1}, \\
\bar{Y}_0 &= \frac{1}{F_0}\sum_{i=1}^{F_0} e_{i,0},
\end{align*}
where $F_j$ denotes the number of participants in arm $j$ who are followed-up, and $A$ denotes the number of participants in the intervention arm who both adhere \emph{and} are follow-up. These sample means have expectations $\mathbb{E}[\bar{Y}_1 | \phi] = \mathbb{E}[\frac{A}{F_1}\mu | \phi] = \phi_a \mu$ and $\mathbb{E}[\bar{Y}_0 | \phi] = 0$. For the variance of the intervention group mean, we have by the law of total variance that
\begin{equation}\label{eqn:int_group_mean}
Var(\bar{Y}_1) = \mathbb{E}[Var(\bar{Y}_1 | F_1, A)] + 
\mathbb{E}[Var(\mathbb{E}[\bar{Y}_1 | F_1, A] | F_1)] + 
Var(\mathbb{E}[\bar{Y}_1 | F_1]).
\end{equation}
Taking each of these terms in turn:
$$
\mathbb{E}[Var(\bar{Y}_1 | F_1, A)] = \mathbb{E} \left[ \frac{\sigma^2}{F_1} \right] = \frac{2\sigma^2}{\phi_f \mathbb{E}[N]},
$$
where $\mathbb{E}[N]$ is the expected number of participants recruited to the trial and randomised equally between arms. Denoting the number of eligible participants who consent by $C$, then,
$$
\mathbb{E}[N] = \mathbb{E}[N ~|~ C < n_t] Pr(C < n_t) + \mathbb{E}[N ~|~ C \geq n_t] Pr(C \geq n_t).
$$
Note that $C$ follows a binomial distribution with size $n_e$ and probability $\phi_r$. Because recruitment will stop once the target has been reached, $\mathbb{E}[N ~|~ C \geq n_t] = n_t$. We also have
$$
\mathbb{E}[N ~|~ C < n_t] = \frac{\sum_{k=0}^{n_t-1} k{n_e \choose k} \phi_r^k (1-\phi_r)^{n_e - k} } {Pr(C < n_t)},
$$
and so 
$$
\mathbb{E}[N] = \sum_{k=0}^{n_t-1} k{n_e \choose k} \phi_r^k (1-\phi_r)^{n_e - k} + n_t Pr(C \geq n_t).
$$
Returning to equation \ref{eqn:int_group_mean}, the second term is 
\begin{align*}
\mathbb{E} \left[ Var(\mathbb{E}[\bar{Y}_1 | F_1, A] ~|~ F_1) \right] &= \mathbb{E} \left[ Var \left(\frac{A\mu}{F_1} ~|~ F_1 \right) \right] \\
&= \mathbb{E} \left[\frac{\mu^2}{F_1^2} Var(A ~|~ F_1) \right] \\
&= \mathbb{E} \left[ \frac{\mu^2}{F_1^2} F_1 \phi_a (1-\phi_a) \right] \\
&= \frac{2\mu^2}{\phi_f \mathbb{E}[N]} \phi_a (1-\phi_a).
\end{align*}
Finally,
\begin{align*}
Var(\mathbb{E}[\bar{Y}_1 | F_1]) &= Var \left( \mathbb{E} \left[ \frac{A \mu}{F_1} ~|~ F_1 \right] \right) \\
&= Var \left( \mathbb{E} \left[ A ~|~ F_1 \right] \frac{\mu}{F_1} \right) \\
&= Var \left( F_1 \phi_a \frac{\mu}{F_1} \right) \\
&= 0.
\end{align*}
This then gives
$$
Var(\bar{Y}_1) = \frac{2\sigma^2}{\phi_f \mathbb{E}[N]} + \frac{2\mu^2}{\phi_f \mathbb{E}[N]} \phi_a (1-\phi_a).
$$

The variance of the control group sample mean is
\begin{align*}
Var(\bar{Y}_0) &= \mathbb{E}[Var(\bar{Y}_0 | F_0)] + Var(\mathbb{E}[\bar{Y}_0 | F_0]) \\
&= \mathbb{E} \left[ \frac{\sigma^2}{F_0} \right] + 0\\
&= \frac{2\sigma^2}{\phi_f \mathbb{E}[N]}.
\end{align*}

The power of the trial can then be obtained by substituting the expectation and variance of the sample means into the usual formula (which, again, assumes they are normally distributed):
\begin{align*}
g(\phi) &= \Phi \left(\frac{\mathbb{E}[\bar{Y}_1 - \bar{Y}_0]}{\sqrt{Var(\bar{Y}_1 - \bar{Y}_0)}} - z_{1-\alpha} \right) \\
&= \Phi \left( \frac{\phi_a\mu}{\sqrt{ \frac{2\sigma^2}{\phi_f \mathbb{E}[N]} + \frac{2\mu^2}{\phi_f \mathbb{E}[N]} \phi_a (1-\phi_a) + \frac{2\sigma^2}{\phi_f \mathbb{E}[N]}} } - z_{1-\alpha} \right) \\
&= \Phi \left( \frac{ \phi_a\mu \sqrt{\phi_f \mathbb{E}[N]} } {\sqrt{4\sigma^2 + 2\mu^2 \phi_a(1-\phi_a)}} - z_{1-\alpha} \right) \\
&= \Phi \left( x(\phi) - z_{1-\alpha} \right),
\end{align*}

\end{document}